\documentclass[sigconf]{acmart}

\setcopyright{none}

\usepackage{graphicx} % Required for inserting images
\usepackage{tcolorbox}
\tcbset{
     colback=gray!12,
    colframe=gray!40,
    coltitle=black,
    fonttitle=\bfseries,
    title=Position Statement,
    boxrule=0.4pt,
    arc=2pt
    colback=gray!10,
  colframe=gray!40,
  boxrule=0.3pt,
  arc=1pt,
  left=3pt,
  right=3pt,
  top=2pt,
  bottom=2pt,
}
\title{Towards Structured, State-Aware, and Execution-Grounded Reasoning for Software Engineering Agents}
\author{Tse-Hsun (Peter) Chen}

\affiliation{%
  \department{Software PErformance, Analysis, and Reliability (SPEAR) lab}
  \institution{Concordia University}
  \city{Montreal}
  \state{QC}
  \country{Canada}
}
\email{peterc@concordia.ca}

\copyrightyear{2026}
\acmYear{2026}
\setcopyright{cc}
\setcctype{by}
\acmConference[BoatSE '26]{7th International Workshop on Bots and Agents in Software Engineering }{April 12--18, 2026}{Rio de Janeiro, Brazil}
\acmBooktitle{7th International Workshop on Bots and Agents in Software Engineering (BoatSE '26), April 12--18, 2026, Rio de Janeiro, Brazil}
\acmPrice{}
\acmDOI{10.1145/3786161.3788456}
\acmISBN{979-8-4007-2393-3/2026/04}

\begin{document}
\begin{abstract}
Software Engineering (SE) agents have shown promising abilities in supporting various SE tasks. Current SE agents remain fundamentally reactive, making decisions mainly based on conversation history and the most recent response. However, this reactive design provides no explicit structure or persistent state within the agent's memory, making long-horizon reasoning challenging. As a result, SE agents struggle to maintain a coherent understanding across reasoning steps, adapt their hypotheses as new evidence emerges, or incorporate execution feedback into the mental reasoning model of the system state.  

In this position paper, we argue that, to further advance SE agents, we need to move beyond reactive behavior toward a structured, state-aware, and execution-grounded reasoning. 
We outline how explicit structure, persistent and evolving state, and the integration of execution-grounded feedback can help SE agents perform more coherent and reliable reasoning in long-horizon tasks. We also provide an initial roadmap for developing next-generation SE agents that can more effectively perform real-world tasks.

%We outline how explicit intermediate representations, persistent and evolving state, and principled use of execution feedback can form a foundation for more reliable and coherent agent reasoning. Our roadmap highlights key research directions for developing next-generation SE agents capable of supporting complex, multi-step workflows in real-world software engineering tasks.
%The agents make decisions based on the most recent response, lack persistent structure, and interpret execution feedback as unstructured text. These limitations prevent agents from maintaining coherent long-horizon reasoning, updating hypotheses systematically, or integrating runtime evidence into a stable mental model of the evolving software system.

\end{abstract}
\maketitle

%Recent studies, such as SWE-Agent~\cite{yang2024sweagent} and AutoCodeRover~\cite{autocoderover}, demonstrate that LLM agents can autonomously navigate repositories, edit code, and show great results on benchmarks such as SWE-bench~\cite{jimenez2024swebench}. 
%However, 
%Software engineering tasks unfold as sequences of interdependent reasoning steps. Each action—running a test, inspecting a failure, applying a patch, adjusting a configuration—introduces new information that must be interpreted relative to what is already known. Studies of debugging, root-cause analysis, and automated program improvement consistently show that developers (and tools attempting to emulate them) must iteratively update their understanding of the system as new evidence appears~\cite{LLMRCA,autocoderover,agent4se-survey}. What makes SE tasks uniquely challenging is not merely that they are multi-step, but that \emph{each} step reshapes the working model of the system.

\section{Introduction}

Developers are rapidly adopting large language model (LLM) agents to assist with a variety of software engineering activities. Beyond simple code completion, recent work shows that Software Engineering (SE) agents can invoke tools, analyze execution feedback, and iteratively update code or configurations~\cite{agentless,yang2024sweagent,autocoderover, agentcoder,soen101}. Emerging agentic SE development platforms such as GitHub Copilot, OpenAI’s GPT-5.1-Codex, Anthropic’s Claude Code, and TRAE~\cite{githubcopilot,gpt51codex,claudecode,trae} also show a trend toward integrating agent-driven workflows into end-to-end software development.

Despite significant progress, current SE agents still operate reactively. Their decisions are based on the most recent prompt and conversation history~\cite{meng2025empiricalstudyllmbasedagents, yao2023react, NEURIPS2023_1b44b878}, with no structured representation or evolving states in the agent memory. Moreover, execution feedback is often incorporated in an ad-hoc manner rather than as an integrated part of the agent's reasoning. 
These limitations make long-horizon planning inherently challenging. Without a persistent structure or state, agents can lose track of their objectives, misinterpret environmental changes, and produce actions that conflict with earlier steps. The absence of stateful connections across tool outputs and reasoning results further prevents agents from forming a coherent understanding of how the system evolves as they act.

In contrast, \textit{real-world development requires structured and stateful reasoning that evolves based on the system state and execution feedback}. 
Developers do not solve tasks step-by-step in isolation. Developers gradually build their \textit{mental model} based on code structure, dependencies, runtime behavior, and their interactions. They create initial hypotheses to reason about the root cause of a failure or the impact of new code changes, and refine the hypotheses after receiving feedback from compilers, test results, debuggers, or logs~\cite{10.1109/TSE.2004.101}. 
Every piece of feedback, whether expected or unexpected, represents a state transition that needs to be integrated or updated into the ongoing reasoning. 
\textbf{This evolving mental model guides subsequent decisions, such as what to inspect next, how a change may propagate, or whether to form a completely new hypothesis}.
%Every feedback helps developers updates their mental model of the system state, guiding them to make subsequent decisions such as where to inspect next, which edits to apply, how broadly a change should propagate, or whether they need to form a completely new hypothesis. 

Current LLM-based agents cannot support structured, execution-grounded reasoning, as all reasoning and execution feedback are represented as unstructured text. 
Without concrete, well-defined structures, agents struggle to capture how system states transition over time or to leverage the implicit conditions that condition those states. 
Although execution feedback provides evidence of how software states change, current agents lack the necessary representation to incorporate this information into a coherent, constantly evolving behavioral model of the system. 
Thus, the agents must iteratively reconstruct their understanding from an unstructured, noisy prompt history, leading to inconsistent, less reliable multi-step reasoning.

In this paper, we argue that \textbf{advancing SE agents requires rethinking the foundations of their reasoning processes}. We believe that effective agent reasoning must incorporate explicit structure, behavior state, and execution-grounded updates. These capabilities mirror how human developers construct and refine agents' mental models to understand the software systems over time. 

\noindent \textbf{Paper Organization}. Section~\ref{mismatch} discusses the current limitation in SE agents. Section~\ref{roadmap} presents the future roadmap for SE agents. Section~\ref{conclusion} concludes the paper.

%%%%%%%%%%%%%%%%%%%%%%%%
%The gap between human-like reasoning and current LLM-based agent behavior becomes most visible when agents are embedded in real development workflows. 

\section{The Fundamental Mismatch in SE Agent Reasoning}
\label{mismatch}

In this section, we discuss the limitations of existing reactive SE agents. We use \textit{\textbf{agent state}} to refer to the agent’s explicit persistent representation of its current understanding of the system (e.g., inferred failure points, invariants, or dependency relations). We use \textit{\textbf{hypothesis}} to refer to any provisional assumption or expectation the agent forms during reasoning (e.g., anticipated behavior after a code change). We use \textit{\textbf{structure}} to describe the organization of this state. The structure describes how hypotheses, dependencies, and observations are stored, updated, and related. 
We distinguish between system state, which refers to the software under analysis's actual runtime or code-level state, and agent state, which captures the agent's current understanding of the system. 
Importantly, we do not treat states as raw conversational memory or retrieved text, but as a structured representation that can be revised and validated. These definitions help clarify the limitations of current reactive agents, which lack explicit mechanisms to maintain or evolve such structured states across reasoning steps.

\subsection{Why SE Tasks Expose the Limits of Reactive Agents}

Recent research finds that the reasoning abilities of these agents degrade as interactions grow longer~\cite{understanding2025agents,agent4se-survey}.  
As interactions grow longer, agents often lose track of prior reasoning, generate inconsistent results, or overfit to the most recent output~\cite{ordermatters}. 
These behaviors suggest that current reactive agent designs lack the mechanisms needed to sustain coherent reasoning over extended SE workflows.

However, software engineering tasks often require long chains of reasoning that integrate evidence across multiple steps. 
Developers must iteratively update their understanding of the system (i.e., system \textit{state}) and the \textit{hypotheses} that guide their reasoning as new evidence appears. More importantly, each step in a task reshapes the system's states and affects the internal model of the entire process. 

Recent agent research has explored various forms of memory to extend context, including retrieval-based memory, vector-store memories, and long-term conversation summaries~\cite{meng2025empiricalstudyllmbasedagents, chhikara2025mem0buildingproductionreadyai, ordermatters, xu2025amem}. These approaches primarily focus on information retention and recall, enabling agents to retrieve past observations. In contrast, our notion of state-aware reasoning focuses on maintaining explicit, structured representations of the agent’s current reasoning commitments, such as hypotheses, invariants, dependencies, and pre- and post-conditions.

This long-horizon reasoning process creates three main challenges: 

\begin{itemize}
    \item \textbf{Historical coherence:} agent decisions must remain consistent with prior insights, or we need to explicitly revise or remove them when they are no longer valid.
    \item \textbf{Interpretive stability:} new evidence must be understood in the context of existing beliefs about the system, rather than in isolation.
    \item \textbf{Pre–post reasoning consistency:} agents must maintain hypotheses about the system’s behavior before and after each action, and systematically update or correct these hypotheses as new feedback arrives.

\end{itemize}

\begin{tcolorbox}We believe addressing these main challenges requires innovative research solutions that \textit{\textbf{maintain and evolve structured agent states across reasoning steps}}. Without such structured memory representations and guidance, agents may exhibit inaccurate reasoning, reasoning drift, and misalignment over the course of a long reasoning process. 
\end{tcolorbox}

\subsection{How the Reactive Paradigm Fails in Practice}
\begin{figure}[t]
    \centering
    \includegraphics[width=1\linewidth]{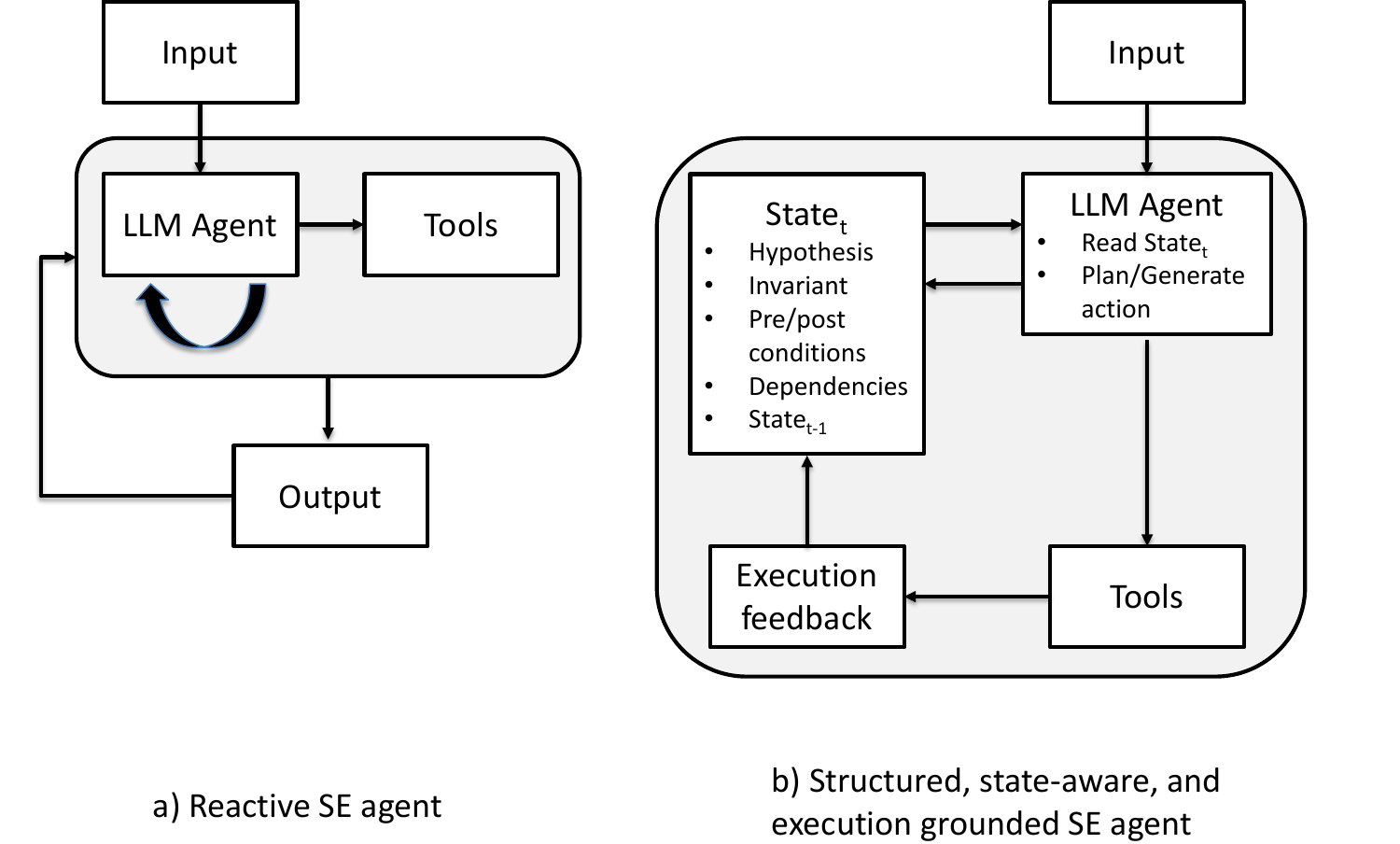}
    \caption{An illustration of the difference between reactive SE agents and structured, state-aware, execution-grounded agents. The example highlights the explicit persistence and evolution of agent state across actions.}
    \label{fig:state-aware-agent}
\end{figure}

Since current SE agents do not explicitly maintain structured reasoning states or track their intermediate hypotheses, they need to repeatedly reconstruct their understanding from the entire prompt history. Figure~\ref{fig:state-aware-agent} illustrates the difference between reactive, and structured, state-aware, and execution-grounded SE agents. Empirical analyses of agent trajectories show that this reconstruction-based process produces several recurring failure patterns in long-horizon SE tasks~\cite{understanding2025agents, zhang2025which,ordermatters}.

\paragraph{\textbf{Inconsistent reasoning reconstructions}.}% violations of historical coherence
Without stable intermediate representations, agents may generate inconsistent explanations or plans across steps. Prior evaluations on benchmarks such as SWE-bench show that repeated runs, even under identical initial conditions, frequently lead to decisions that contradict earlier reasoning~\cite{understanding2025agents, 2026agentsama}. We believe these inconsistencies are \textit{more prevalent} because each reconstruction attends to different parts of the available context, making it impossible to maintain a coherent logical link to earlier insights. 

\paragraph{\textbf{Forgotten assumptions and broken reasoning process}.}
The ``\textit{assumptions}'' or ``\textit{hypotheses}'' that the SE agents made across actions in the reasoning process are not tracked and maintained in agent memory. These assumptions, such as suspected failure points, invariants, or expected behavior after code changes, are \textit{important hypotheses that affect reasoning decisions}. 
Bouzenia et al.~\cite{understanding2025agents} show that agents often contradict earlier self-generated constraints or reintroduce issues they had resolved previously. Without ways to store or revise hypotheses, agents cannot preserve or update the rationale that motivates their actions.

\paragraph{\textbf{Interpreting execution feedback in isolation}.}
Currently, execution feedback, such as test execution, runtime traces, and logs, is provided to agents in unstructured text format. 
Because agents lack clear records of what they assumed in earlier steps, they often interpret this feedback in isolation rather than as part of an ongoing line of reasoning. 
Prior work~\cite{LLMRCA,TRAIL,ordermatters} shows that this leads agents to focus on the most recent messages or the incorrect part of the system. Since execution feedback is not connected to their earlier reasoning, agents struggle to form a stable view of what the system is doing or how their hypotheses should change.

\paragraph{\textbf{Retrying without knowing where to return}.} When errors or contradictions occur in the reasoning process, many agents perform a reset-and-retry strategy. However, most agents often restart the entire process from the beginning, rather than identifying a specific step where the reasoning went off course~\cite{zhang2025which,understanding2025agents}. Localizing the specific problematic step can be challenging without a proper agent memory structure to represent the actions. A complete restart also incurs additional overhead, produces inconsistent output, or may even cause the agent to repeat the same failure, as all prior reasoning was removed.

The above-mentioned failure patterns are model-agnostic and tied to the reactive nature of the current SE agent design. 
We believe addressing these limitations requires rethinking how agents represent, update, and rely on intermediate reasoning states.

\section{Rethinking SE Agent Reasoning: From Reactive to Structured}
\label{roadmap}

In this section, we outline an initial roadmap for moving from reactive SE agents to structured, state-aware, and execution-grounded reasoning.

\paragraph{\textbf{From Conversation Histories to Explicit Intermediate Representations}.}

To support long-horizon tasks, instead of directly using the conversation history as agent memory, agents should leverage \emph{explicit intermediate representations} that stores:

\begin{itemize}
    \item the current understanding of the relevant code elements and their dependencies,
    \item assumptions and hypotheses formed during earlier steps,
    \item expected behavior or invariants of the actions, and
    \item tentative or alternative hypotheses that may be further explored.
\end{itemize}

Our recent study~\cite{2026agentsama} shows that representing agent actions using a finite-state machine abstraction with explicit pre- and post-conditions (e.g., similar to forming hypotheses before making actions) leads to more stable and accurate multi-step reasoning. 
Such structured representations help separate completed actions from unresolved hypotheses, allow systematic refinement of earlier assumptions, and avoid interpreting long conversational histories.

\paragraph{\textbf{Reasoning as an Evolving State Rather Than Repeated Reconstruction}}

Developers maintain and continuously refine a mental model of the task that they are solving. To emulate this behavior, agents should treat reasoning as the addition, deletion, or evolution of an internal state rather than a sequence of independent actions. 
The agent should update the state, such as revise the hypothesis, adjust expected behaviour, or include new evidence, after each action or tool invocation. 

Maintaining the states keeps the reasoning process cleaner and more structured. This reduces the noise an LLM receives and avoids the need to restart the entire process when an error occurs. A structured representation also makes it easier for the agent to self-reflect and pinpoint where an incorrect assumption was introduced in the reasoning process.

\paragraph{\textbf{Connecting Execution Feedback to Structured Updates}}

Existing agents often treat execution feedback as text appended to the prompt~\cite{agentless,soen101, autocoderover,yang2024sweagent,agentcoder}. However, software engineering tasks require interpreting feedback based on prior assumptions about system behavior and code structure. A structured agent must therefore:

\begin{itemize}
    \item map execution feedback to the corresponding assumptions or hypotheses (e.g., a failing test indicating that an assumed invariant about the system’s behavior does not hold),
    \item identify which components of the internal state are affected and how,
    \item revise any assumptions or planned next steps that depend on the feedback, and
    \item update hypotheses about how the system is expected to behave in subsequent steps. 
\end{itemize}

Treating execution feedback as structured evidence allows the agent to incorporate new information into its internal state, rather than responding to each observation independently. We believe this gives agents a more stable and coherent understanding of the system’s behavior.

Overall, rethinking SE agent reasoning calls for an evolution from reactive behavior toward structured, reasoning-grounded processes. 
These directions point to a potential roadmap for a new generation of SE agents capable of maintaining coherent, state-driven reasoning across long-horizon tasks.

\section{Conclusion}
\label{conclusion}

Current Software Engineering (SE) agents remain fundamentally reactive, making decisions mainly based on the most recent response. However, we argue that effective agent reasoning requires explicit structure, persistent state, and integration of execution-grounded feedback. 
These principles are critical for agents to maintain coherent reasoning across steps, adapt their understanding as new evidence appears, and better support complex workflows in real-world SE tasks.

Our roadmap outlines an initial direction for developing more reliable and capable SE agents. Moving toward structured, state-aware, and execution-grounded reasoning will require new agent memory representations and better integration with analysis and runtime tools to improve the need for long-horizon reasoning in SE tasks.

%The limitations of current reactive Software Engineering (SE) agents highlight the need for reasoning models that operate over explicit structure, evolving state, and grounded feedback. Our position paper argues that effective SE agents must support three core capabilities.

%First, \textbf{revisability}: assumptions and intermediate conclusions must be represented in a form that can be updated or discarded as new evidence appears, without requiring the agent to restart its reasoning process. Second, \textbf{inspectability}: intermediate states should be accessible both to the agent and to external tools or developers, enabling clearer understanding of how decisions were reached and where reasoning may have diverged. Third, \textbf{tool connectivity}: outputs from compilers, tests, static analyses, and profilers should map directly onto elements of the agent’s internal state, creating consistent and interpretable state transitions.

%Together, these capabilities define the foundation for structured, state-aware, and execution-grounded SE agents. By moving beyond prompt-driven behavior toward explicit representations and controlled state evolution, future agents can reason more reliably across complex software engineering workflows and form the basis for a new generation of trustworthy, verifiable agentic systems.

\balance
\bibliographystyle{ACM-Reference-Format}
\bibliography{references}

\end{document}